\documentclass{article}
\usepackage{graphicx}
\usepackage{amsmath, amsfonts, amssymb}
\usepackage{booktabs}
\usepackage{fullpage}
\usepackage[square,sort,comma,numbers]{natbib}
\usepackage{setspace}

\usepackage{color}

\newcommand{\EE}{{\mathsf E}}
\usepackage[normalem]{ulem}
\usepackage{hyperref}
\usepackage{array, caption, floatrow, tabularx, makecell, booktabs}

\title{Discussing Diminishing Returns:\\A New Scoring System for Powerlifting}
\author{Jean C. Peyen \thanks{University of Dundee, \href{mailto:jpeyen001@dundee.ac.uk}{jpeyen001@dundee.ac.uk}} \and Ruheyan Nuermaimaiti\thanks{University of Leeds, \href{mailto:r.nuermaimaiti@leeds.ac.uk}{r.nuermaimaiti@leeds.ac.uk}} \and Joshua Cunningham \thanks{Queen's University of Belfast, \href{mailto:jcunningham1429@qub.ac.uk}{jcunningham1429@qub.ac.uk}}}

\begin{document}

\maketitle
\begin{abstract}
Maximal strength increases with body weight, this is why scoring methods have been developed in order to fairly scale powerlifting performances based on athletes’ body weight. The International Powerlifting Federation (IPF) Good Lift (GL) system, introduced in 2020, is a scaling method based on a Von Bertalanffy function. It is specifically tailored to elite powerlifters—defined as those achieving at least 84$\%$ of the current world record in their weight class, according to the IPF's ``Golden Standard". A key assumption of the GL system is a principle of diminishing returns, stating that performance gains decrease as body weight increases. However, data from the Open Powerlifting database reveals the presence of a phase of increasing returns below a certain body weight threshold. To capture this initial phase, we propose a new scoring system based on a logistic function. This approach is designed to establish strength standards for the general population and for athletes not specialised in powerlifting.
\end{abstract}

\noindent\textbf{Keywords:} powerlifting, growth models, diminishing returns, Von Bertalanffy function, logistic function, resampling, kernel density estimate\\

%\section{Introduction}
Powerlifting is a strength sport consisting of three movements: the squat, the bench press, and the deadlift. In a powerlifting competition, an athlete must perform three attempts in each movement. Only the best attempt in each movement is kept, and the total of these three attempts determines the winner of the competition. This paper is using the Open Powerlifting dataset \cite{openpower}. This is a continually updated dataset that compiles competition results from multiple powerlifting federations since as early as 1964.
We only consider unequipped (raw) athletes in the open age division.

The present analysis focuses on the relationship between body weight and total lifted (thus, we only consider athletes who performed at least one valid attempt per movement).

\begin{figure}[h]
\centering
\begin{minipage}{0.48\textwidth}
\includegraphics[width=8.6cm]{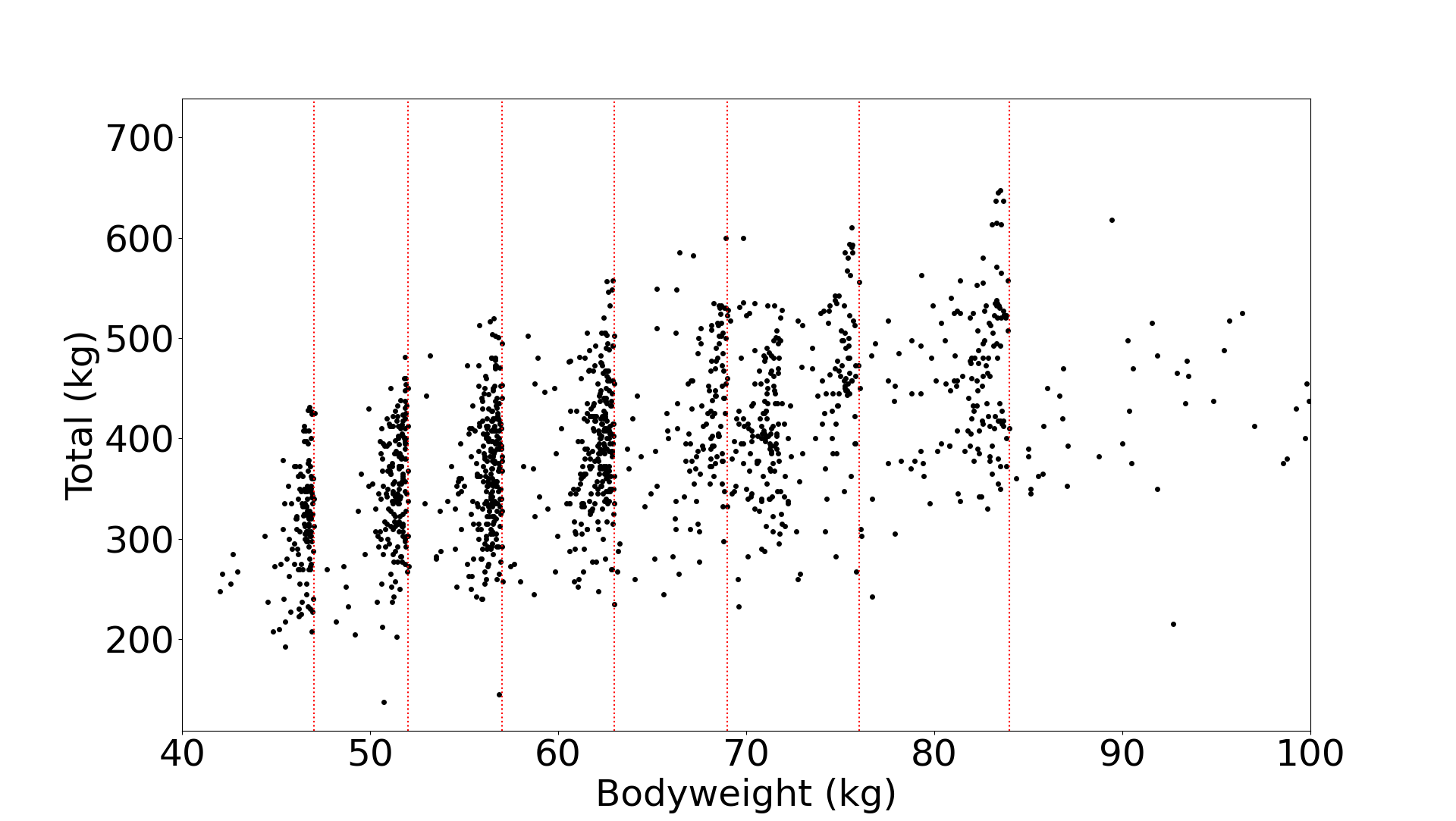}
\centering
\end{minipage}
\begin{minipage}{0.48\textwidth}
\includegraphics[width=8.6cm]{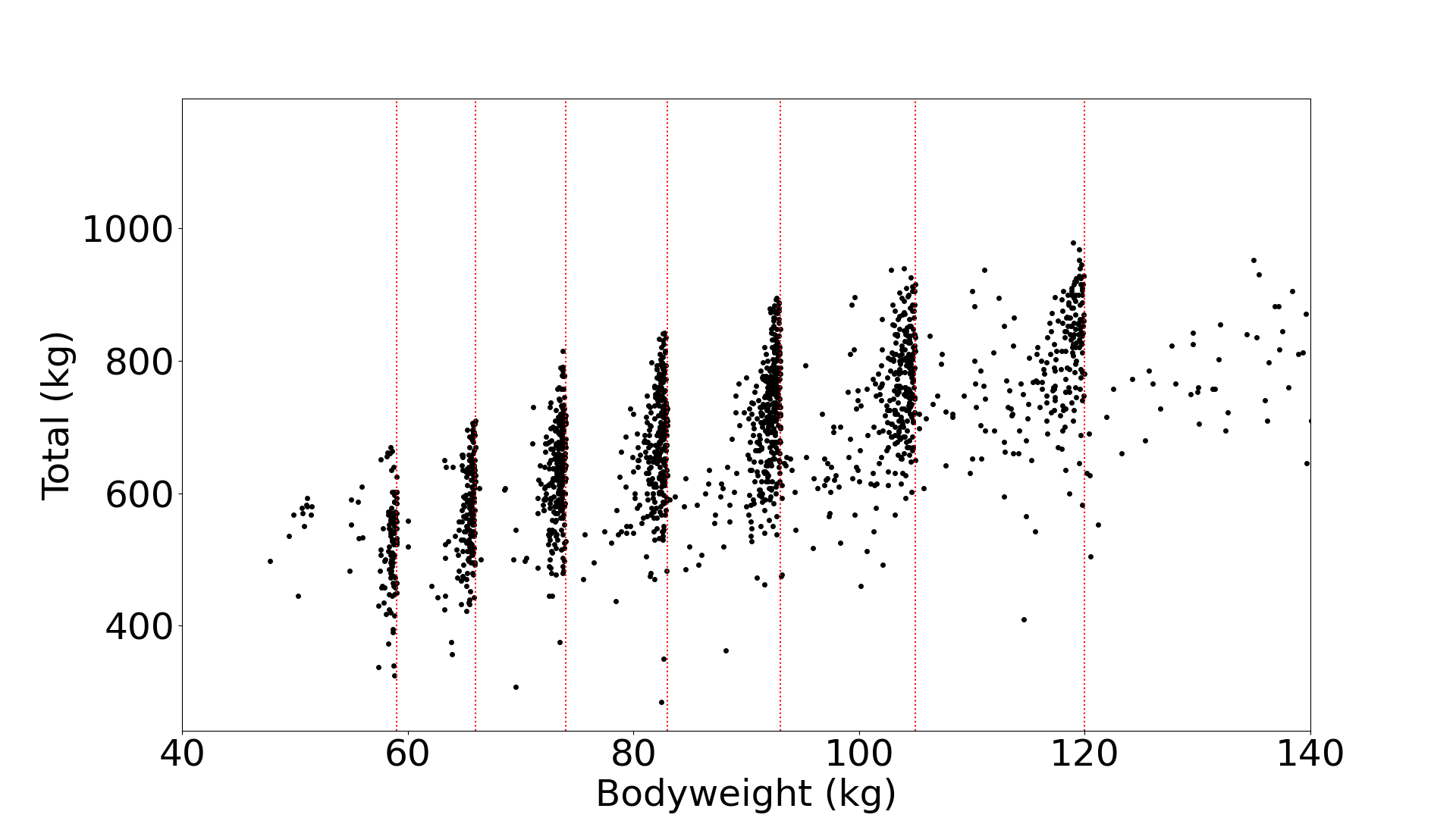}
\centering
\end{minipage}
\caption{Total lifted by Females (left panel) and Males (right panel) in IPF competitions and IPF weight classes (red dotted lines)}
\label{fig:IPF}
\end{figure}

In competitions, the heavier athletes are expected to lift heavier weights, so federations have established weight classes to ensure fair competitions. For example, Table \ref{weightclasses} shows the weight classes on World Games and International Powerlifting Federation (IPF) open division. In order to take advantage of this system, athletes can adopt drastic measures, such as a restrictive diet, or dehydration to enter a weight class which is below their natural body weight while maintaining their muscle mass \cite{Suazo}. The effect of this practice can be observed in Figure \ref{fig:IPF} where we can see an accumulation of the top performances near the upper value of the weight classes. Interestingly, this phenomenon is more prominent in males, indicating that gender-related factors significantly influence the ability to lose weight for an upcoming competition. This is consistent with the fact that testosterone is known to facilitate fat reduction \cite{Corona}, and hormonal fluctuations during the menstrual cycle contribute to variations in fluid retention \cite{Stachenfeld}.

\begin{table}[h!]
\centering
\begin{tabular}{|l|c|c|c|c|}
\hline
\textbf{Category} & \textbf{World Games } & \textbf{IPF Open } & 
\textbf{World Games } & \textbf{IPF Open }\\ 
& \textbf{Females (kg)} &\textbf{Females (kg)} & \textbf{Males (kg)}&\textbf{Males (kg)}\\ \hline
\textbf{Lightweight} & Up to 57.0 & 47.0, 52.0, 57.0& Up to 74.0 & 59.0, 66.0, 74.0  \\ \hline
\textbf{Middleweight} & 57.01 - 63.0 & 63.0 & 74.01 - 93.0 & 83.0, 93.0 \\ \hline
\textbf{Heavyweight} & 63.01 - 72.0 & 69.0, 76.0& 93.01 - 120.0 & 105.0, 120.0 \\ \hline
\textbf{Super Heavyweight} & Over 72.0 & Over 84.0& Over 120.0 & Over 120.0 \\ \hline
\end{tabular}
\caption{World Games and IPF Open Weight Classes for Females}
\label{weightclasses}
\end{table}

To compare the athletes from different weight classes, performances can be scaled according the athletes' body weight. An ideal scoring formula should take the following form
\begin{equation}
	\text{Score } \propto\frac{y}{f(x)},
\end{equation}
where $x$ and $y$ are respectively the body weight and the total of the scored athletes, $f(x)=\EE(y|x)$ is the expected total for a random athlete of the same body weight. Implementing a scoring system can have multiple benefits aside from determining the strongest athlete ``pound for pound". Federations may consider using a scoring system for international qualification events to allow athletes to transition smoothly between weight classes instead of undergoing unhealthy diets. This could also help comparing overall performances of different teams.

%\rncom{more extensive literature review of scores}

In practice, $f$ is a nonlinear function, and various models have been proposed to estimate it. These models can be based on a sufficiently flexible family, such as an ad hoc polynomial family, which fits the data well but has limited explanatory power and discards any notion of parsimony. This is what is done with the Wilks score, which rescales the total by a $5$-th order polynomial \cite{vanderburgh}
\begin{equation}\label{eq:Wilks}
\text{Wilks Score} = \frac{C\cdot y}{a + bx + cx^2 + dx^3 + ex^4 + fx^5}, 
\end{equation}
where $x$ is the athlete's body weight (in kg), $y$ is the athlete's total (in kg) and $f_{\text{Wilks}}(x)=a + bx + cx^2 + dx^3 + ex^4 + fx^5$. 
The coefficients $a,b,c,d,e$ and $f$ are specific to the sex of the lifter. The constant $C$ is set to $500$ in the original version of the Wilks formula and to $600$ in the 2020 version known as Wilks-2. %The author was unable to find the original reference that provides the values of these coefficients; 
The coefficients can be found, for instance, in the IPF models evaluation report \cite{IPFGL}.%\url{https://www.powerlifting.sport/fileadmin/ipf/data/ipf-formula/Models_Evaluation-I-2020.pdf}.

Alternatively, $f$ can be chosen from a family of functions that conforms to some postulates, providing a more interpretable and theoretically grounded model. In 2020, the International Powerlifting Federation (IPF) has implemented the Good Lift (GL) score to replace the Wilks score \cite{IPFGL}. The IPF GL score is defined by
\begin{equation}\label{eq:GL}
\text{IPF Score} = \frac{100\cdot y}{A-B\cdot \exp{}(-Cx)},
\end{equation}
where the coefficients $A,B$ and $C$ are all positive and $f_{\text{IPF GL}}(x)= A-B\cdot \exp{}(-Cx)$. The IPF GL score is justified by the following set of assumptions:
\begin{enumerate}
    \item \textit{With increasing body weight, the absolute lifted weight increases,}
    \item \textit{the relative performance (defined as the total lifted divided by the body weight) decreases,} 
    \item \textit{there is a potential physical limit for the result both for each value of body weight and the absolute weight limit for the entire population.}
\end{enumerate}

\begin{figure}
    \centering
    \includegraphics[width=0.8\linewidth]{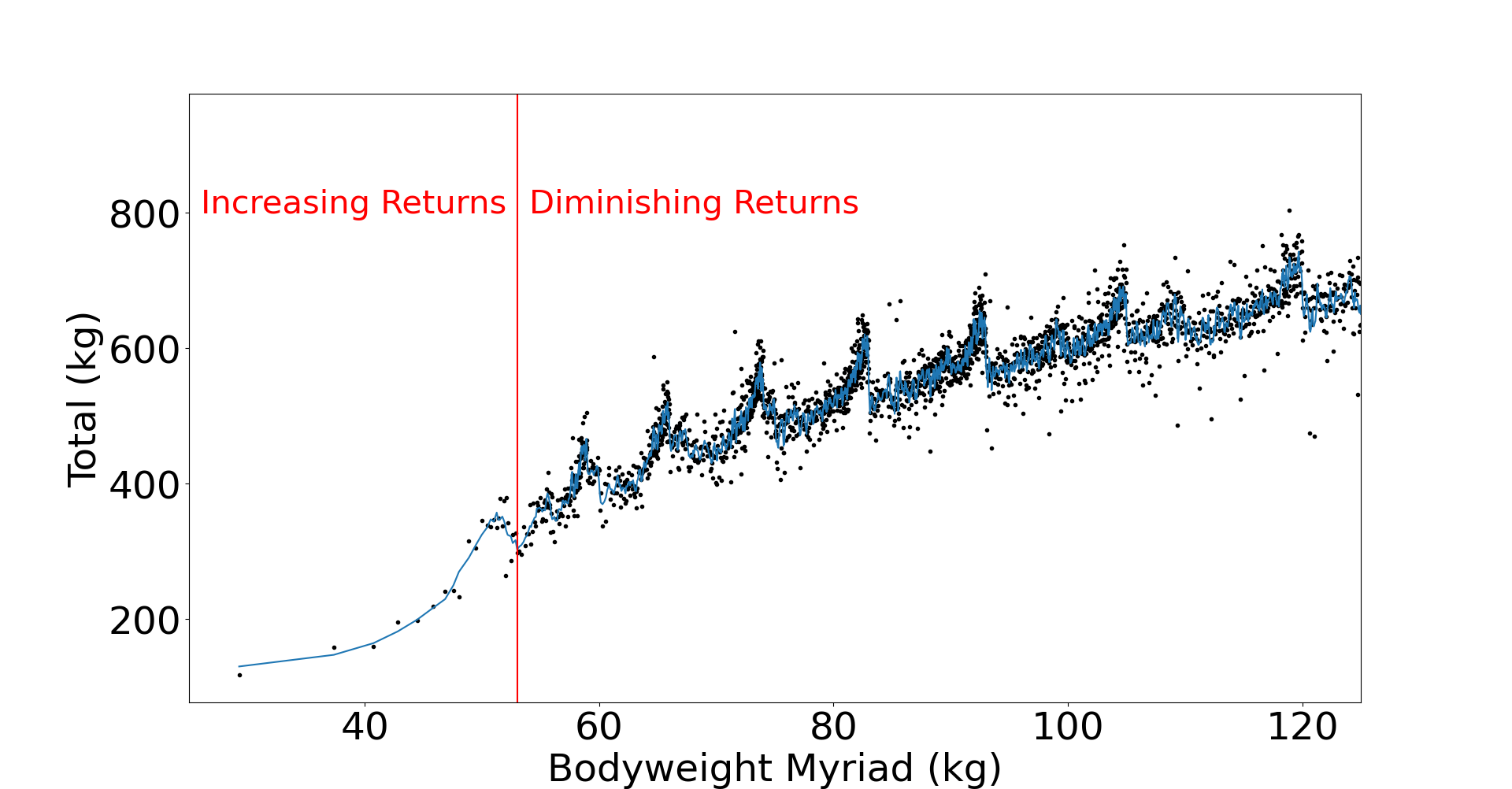}
    \caption{Average total lifted by males in each body weight myriad for the Open Powerlifting dataset}
    \label{fig:dimireturns}
\end{figure}

The second assumption, which can be interpreted as a principle of diminishing returns —see figure \ref{fig:dimireturns}—may be challenged as below a certain body weight, an individual may lack the necessary mass to maintain optimal physiological functions (e.g. of vital organs and passive structures).
This suggests that we should consider a model that incorporates an inflection point below which the relative strength of an athlete would be an increasing function of the body weight. %\rncom{both below and above should be introduced here. From assumptions to here need to be revised.}

\section{Methods}
This section is structured as follows: First, we review the definition of the exponential family used to design the IPF GL score. Next, we review the definition of the logistic function. Finally, we propose to resample the dataset in order to obtain fits that are suitable for athletes with extreme body weight. Resampling is performed using a Kernel Density Estimate (KDE) of the body weight distribution.

\subsection{Von Bertalanffy function (IPF GL score)}\label{sec:GL}
We replicate the type of formula used for the IPF GL score. It is a variant of the Von Bertalanffy function, which can be presented in the following way:
\begin{equation}\label{eq:GL3}
    f(x) = L\left[1-\exp{}(-k(x-x_0))\right]
\end{equation}
It exhibits an initial phase of linear growth, followed by a slowing of the growth rate. It is commonly employed to model the growth of animals as they age \cite{Cailliet}. This behaviour is similar economic growth observed in developed countries, which is characterised by diminishing returns \cite{mankiw}.

We observe that it satisfies the three aforementioned assumptions \cite{IPFGL}. First of all, it is strictly increasing as we have the following:
\begin{equation}
    f' = k\cdot(L-f) > 0.
\end{equation}
It also satisfies a principle of diminishing returns:
\begin{equation}
    f'' = k^2 \cdot (f-L) < 0.
\end{equation}
Finally, $L$ is the limit as the body weight goes to infinity.

\subsection{Logistic function}\label{sec:logistic}
We propose to use a scaling formula based on a logistic function, adjusted in order to meet the condition $f(0)=0$:
\begin{equation}
    f(x) = L\cdot \left(\frac{1}{1 + \exp{}(-k(x - x_0))}-\frac{1}{1+\exp{}(k\cdot x_0)}\right).
\end{equation}
Here, the function $f$ is intended to estimate the total of a random athlete of a given body weight. In the present paper, the values of $L$, $k$ and $x_0$ are adjusted based on the totals recorded for raw athletes in the open division, regardless of their level.
As previously, we can compute the first order derivative:
\begin{equation}
    f'= \frac{k}{L}\cdot \exp{}\left(-k(x-x_0)\right)\cdot f^2
\end{equation}
and the second order derivative:
\begin{equation}
    f'' = -\frac{k}{L}\cdot \exp{}(-k(x-x_0))\cdot f \cdot \left(f-2f'\right).
\end{equation}
This function is characterised by an initial phase of increasing returns, when $0<x<x_0$, followed by a phase of diminishing returns, when $x>x_0$. Of course, we can check that, up to a reparametrisation, this is equivalent to the Von Bertalanffy function when $x$ is large by calculating a first order approximation.

Anecdotally, we should mention that the logisitic function was initially introduced by Verhulst to cap the uncontrolled growth seen in the Malthusian model \cite{verhulst}. The problem that we have here is dual, as we introduce a phase of increasing returns preceding a phase of diminishing returns.

\subsection{KDE resampling}\label{sec:kde}
\begin{figure}[h]
\centering
\begin{minipage}{0.48\textwidth}
\includegraphics[width=8.8cm]{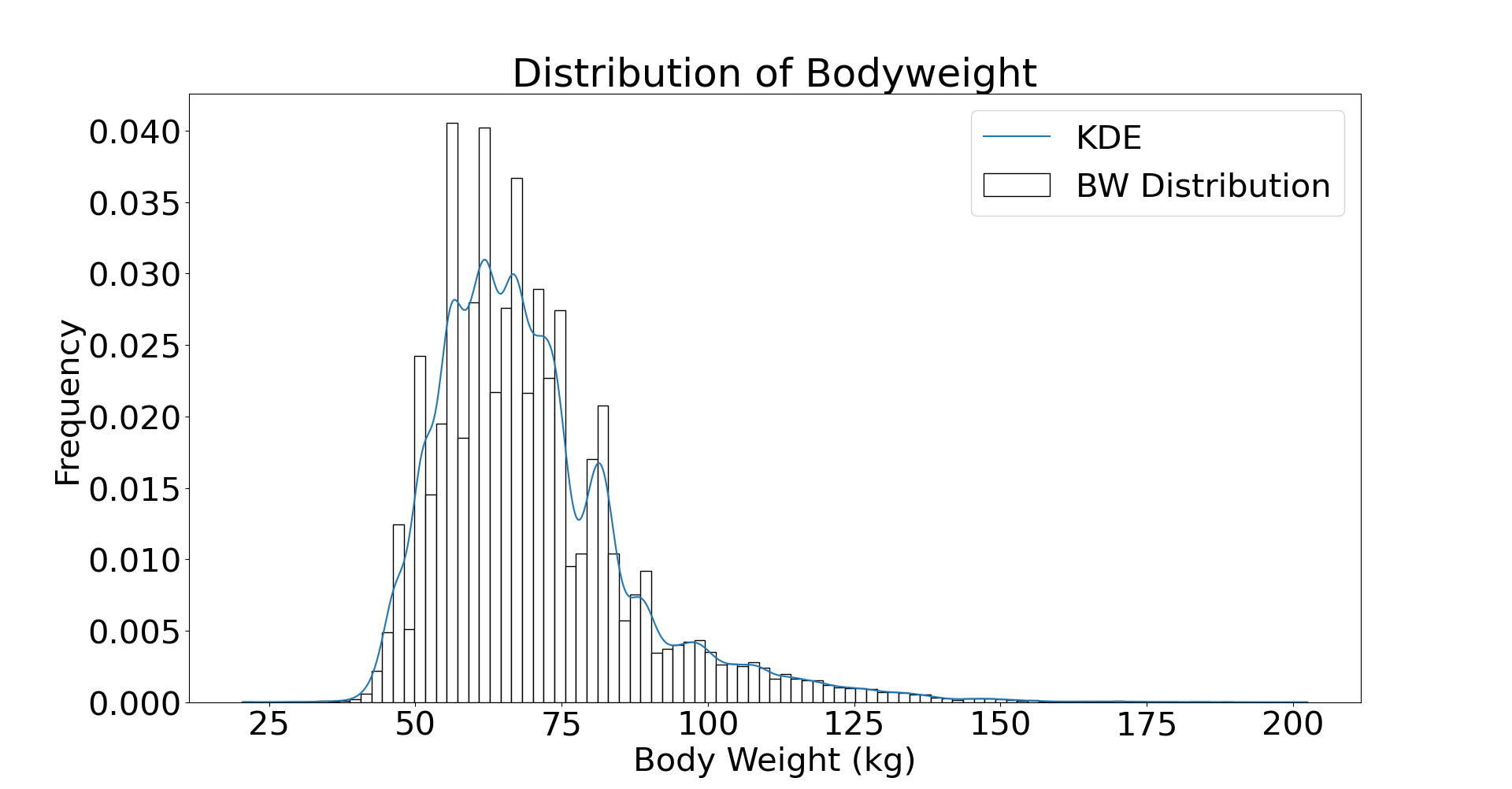}
\centering
\end{minipage}
\begin{minipage}{0.48\textwidth}
\includegraphics[width=8.8cm]{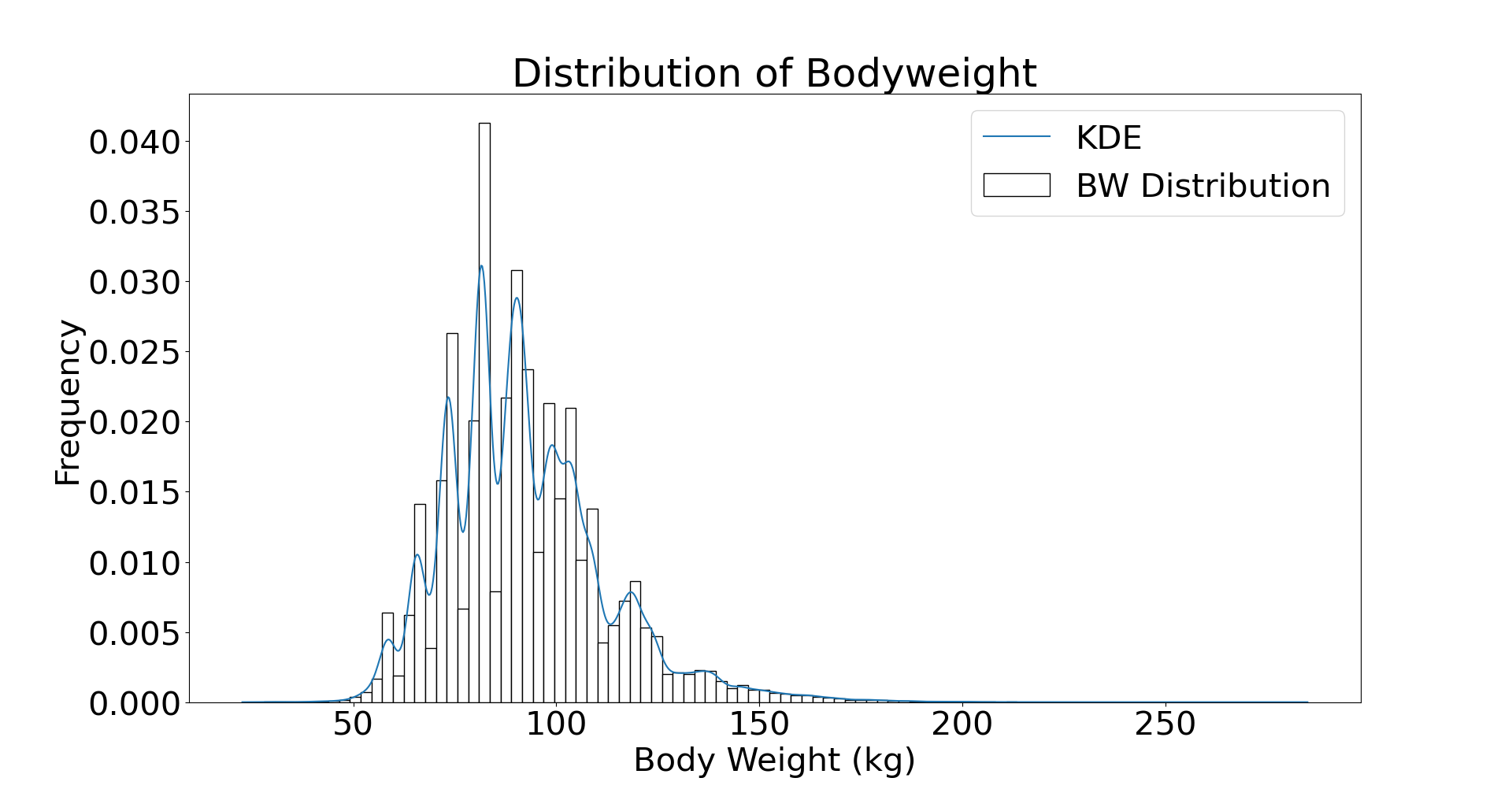}
\centering
\end{minipage}
\caption{Body weight distribution before resampling : females (left panel), males (right panel)}
\label{fig:BWdist}
\end{figure}

As observed in Figure \ref{fig:BWdist}, athletes with extreme body weights are under represented. We use a resampling method in order to have a good fit through the whole range of body weights. We also apply a Gaussian noise in order to obtain more explicit figures. 

An approach to resampling consists in assigning weights to the entries of a dataset $x_1,\cdots ,x_n$ based on an estimation $\hat{g}$ of the density of the body weight distribution. Resampling weights are defined as:
\begin{equation}
    w_i = \frac{1}{\hat{g}(x_i)}.
\end{equation}
Using these weights, $k$ points are resampled from the dataset with the following distribution:
\begin{equation}
    P(x_i) = \frac{w_i}{\sum_{j=1}^n w_j}.
\end{equation}

\begin{figure}[h]
\centering
\begin{minipage}{0.48\textwidth}
\includegraphics[width=8.8cm]{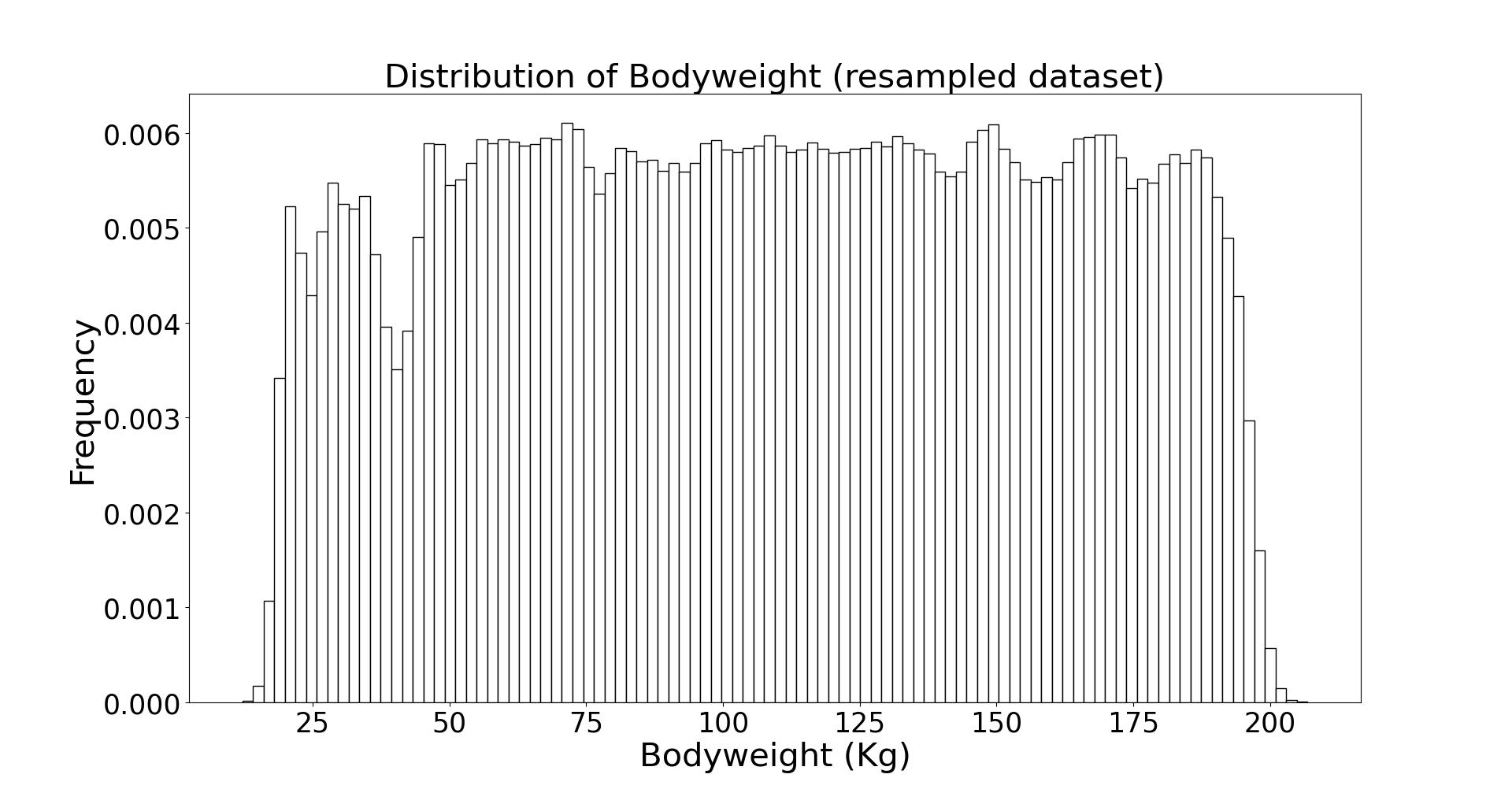}
\centering
\end{minipage}
\begin{minipage}{0.48\textwidth}
\includegraphics[width=8.8cm]{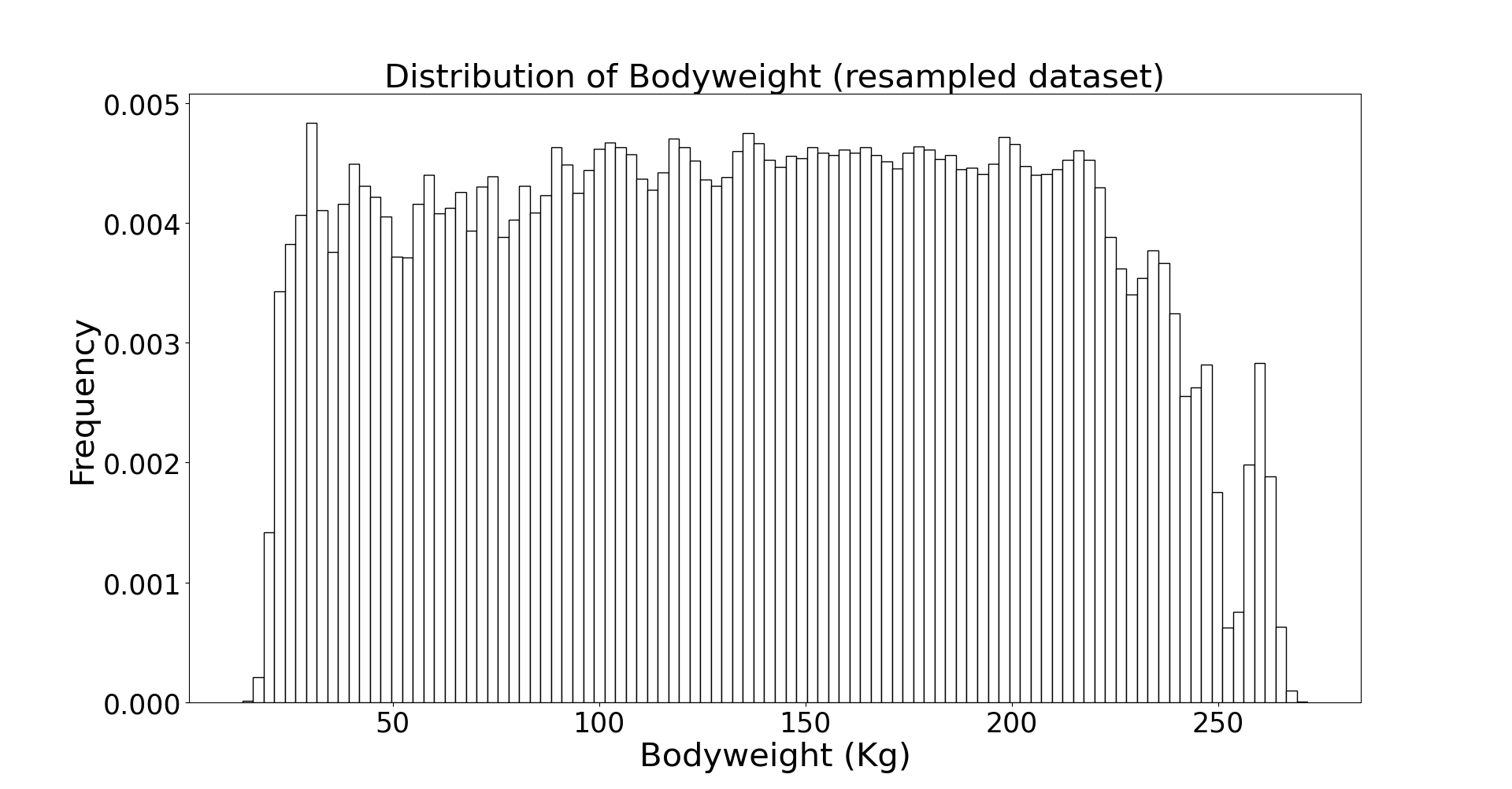}
\centering
\end{minipage}
\caption{Body weight distribution after resampling : females (left panel), males (right panel)}
\label{fig:resampling}
\end{figure}

The body weight distribution has multiple modes, as athletes benefit from being close to the upper limit of their class. In order to account for this, we use a Kernel Density Estimate (KDE):
\begin{equation}
    \hat{g}(x) = \frac{1}{nh}\sum_{i=1}^n K\left(\frac{x_i-x}{h}\right),
\end{equation}
where $K$ is the Gaussian kernel:
\begin{equation}
    K(x) = \frac{1}{\sqrt{2\pi}}\exp{\left(-\frac{x^2}{2}\right)},
\end{equation}
and the parameter $h>0$ is selected with the Scott's rule:
\begin{equation}
    h = n^{-1/5}.
\end{equation}
For more details the reader may refer to \cite[Chap. 3]{silverman}. In \cite{Repo}, the KDE is fitted using the \verb|gaussian_kde| function from the \verb|scipy| package \cite{SciPy}. As we can see in Figure (\ref{fig:resampling}), the resulting dataset has a more uniform distribution.

\newpage
\section{Results}
In this section we summarise the outcomes of our analysis. A code to reproduce this analysis can be found  at \cite{Repo}. The Von Bertalanffy function and the logistic function have initially been fitted to a filtered version of the Open Powerlifting dataset. Then these functions have been fitted to a resampled dataset of size $100,000$. 

\subsection{Fitted models and examples of scores}
\begin{figure}[h]
\hspace{-3pc}
\centering
\begin{minipage}{0.48\textwidth}
\includegraphics[width=8.8cm]{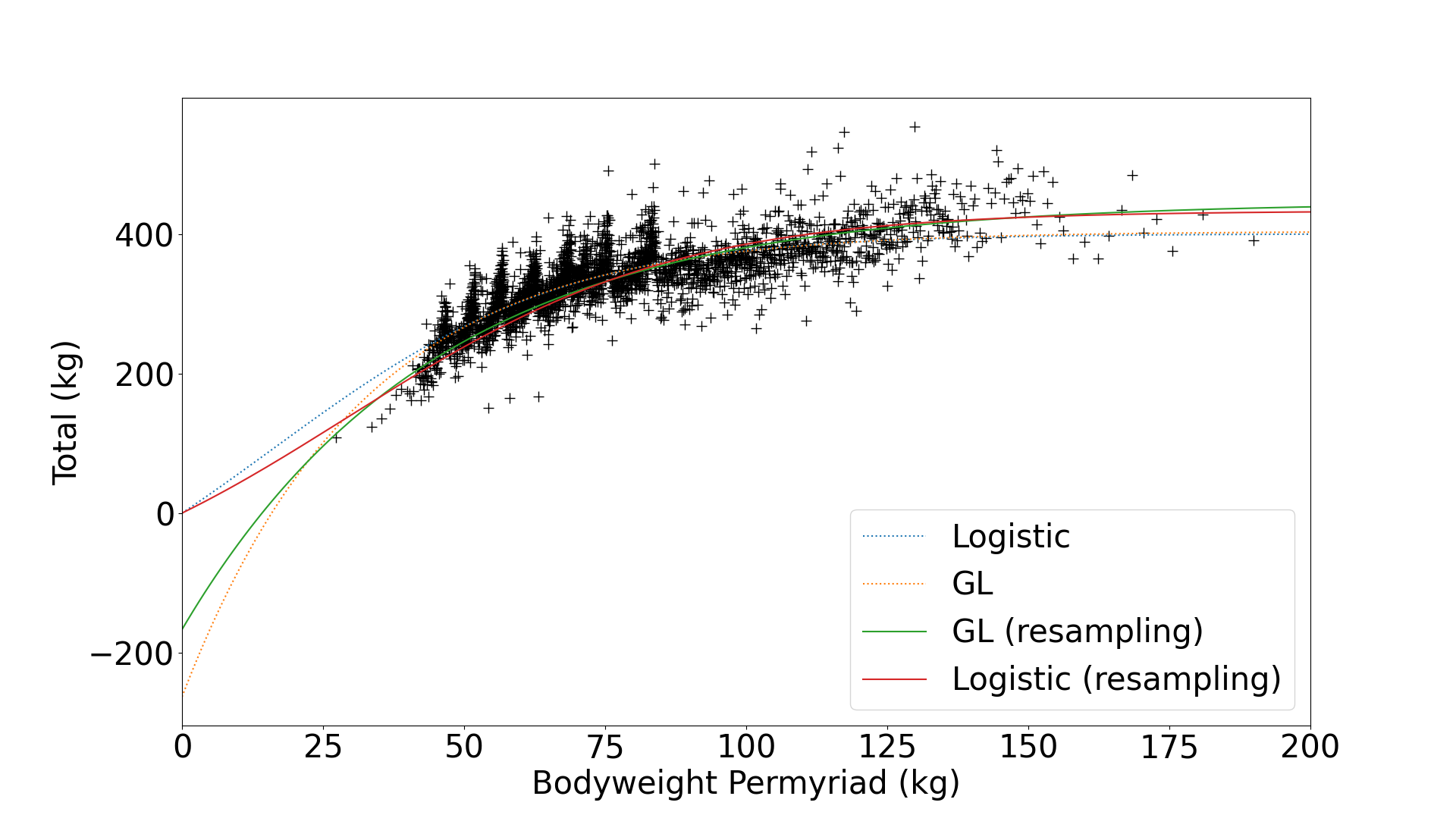}
\centering
\end{minipage}
\begin{minipage}{0.48\textwidth}
\includegraphics[width=8.8cm]{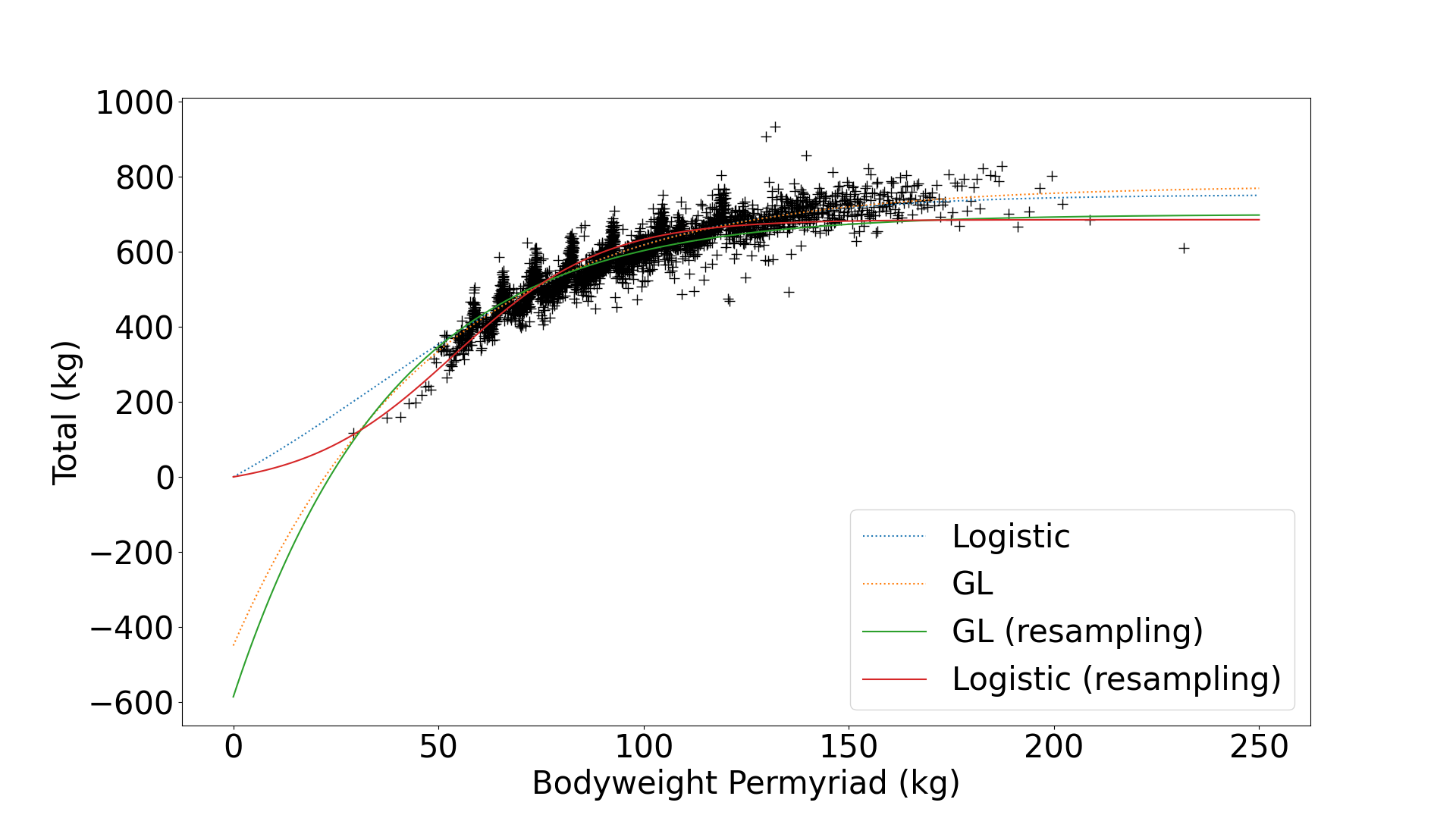}
\centering
\end{minipage}
\caption{Fitted models and myriad from original dataset: females (left panel), males (right panel)}
\label{fig:TotnoReMyr}
\end{figure}

\begin{figure}[h!]
\hspace{-3pc}
\centering
\begin{minipage}{0.48\textwidth}
\includegraphics[width=8.8cm]{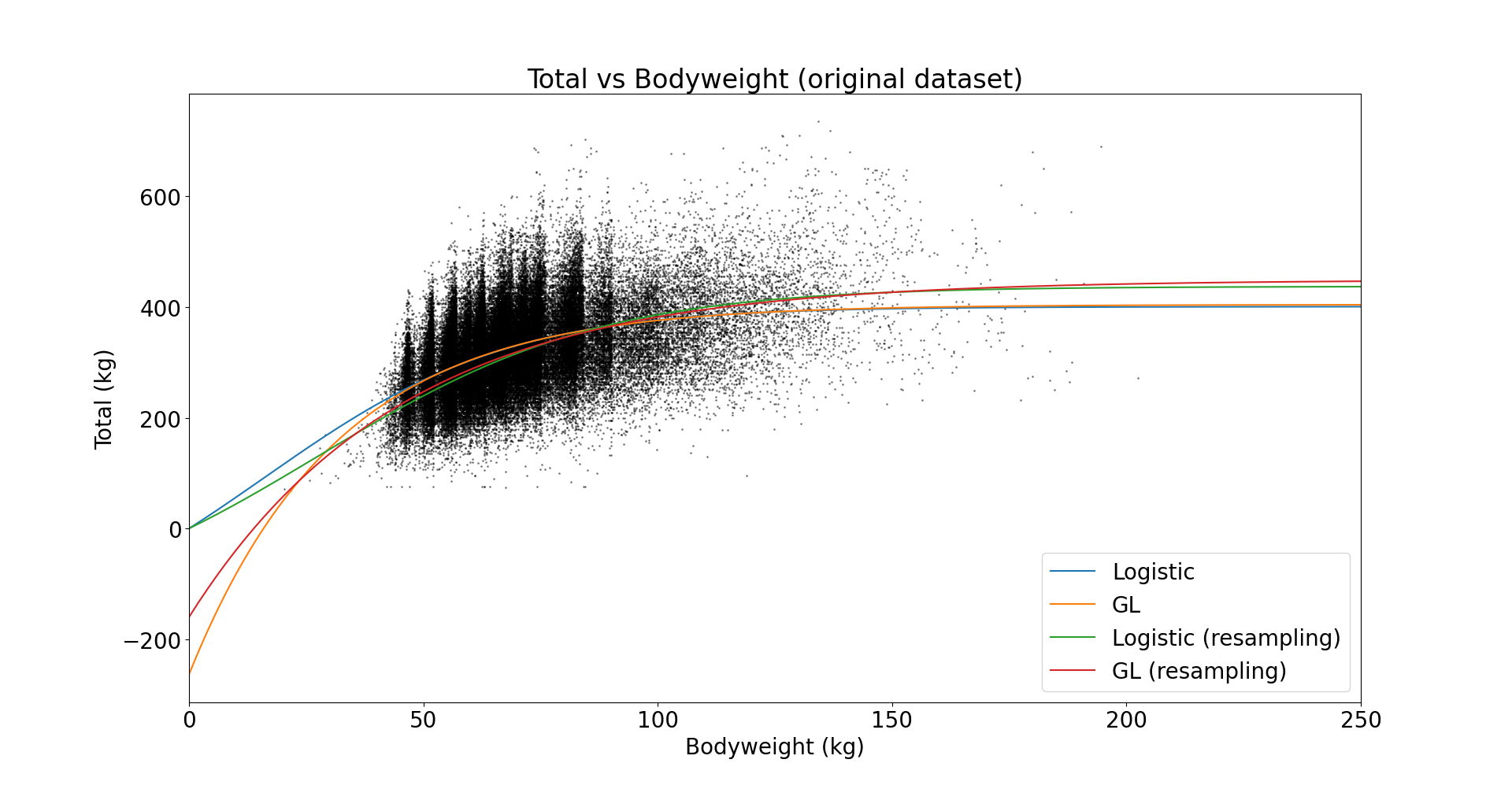}
\centering
\end{minipage}
\begin{minipage}{0.48\textwidth}
\includegraphics[width=8.8cm]{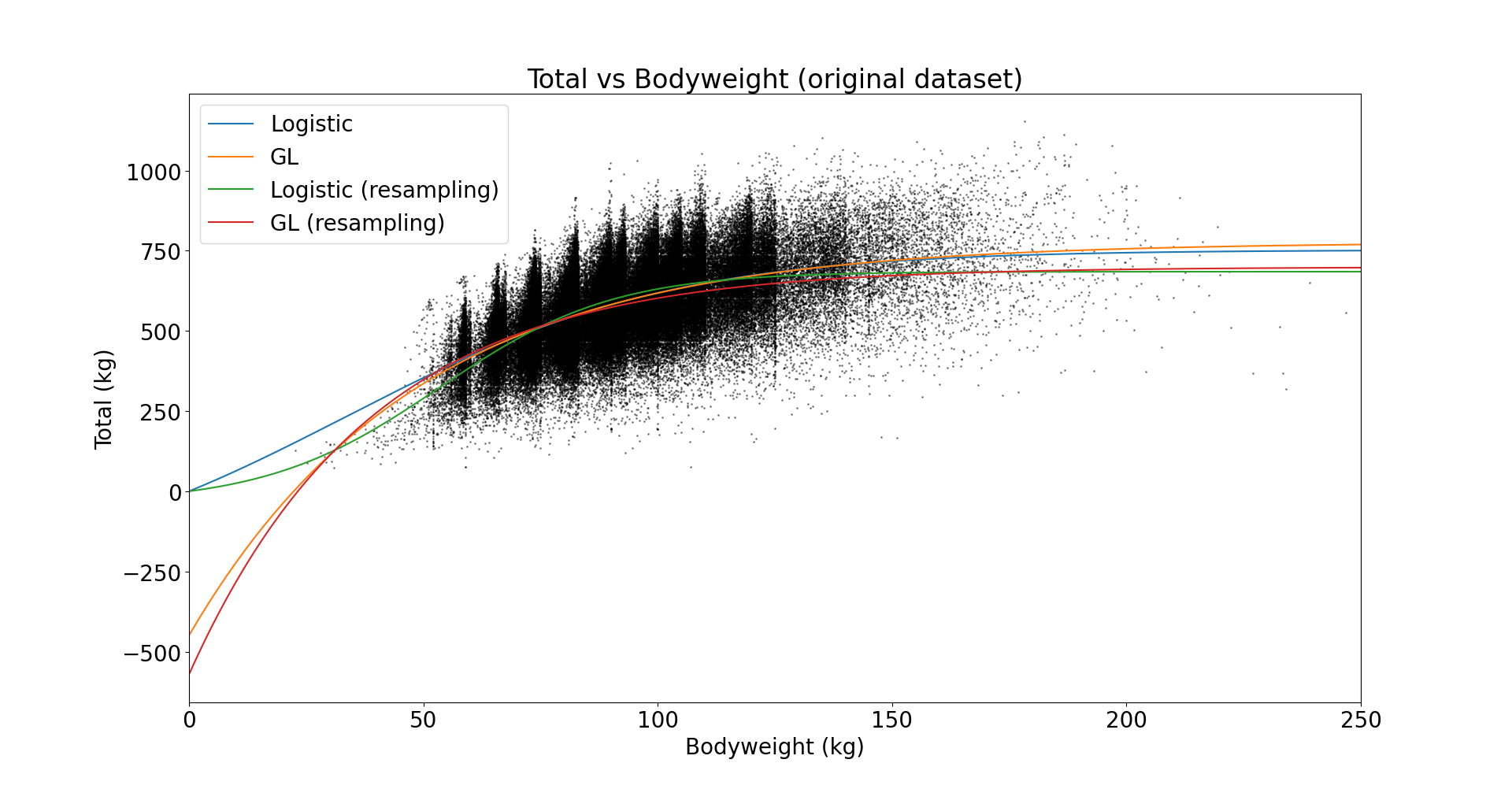}
\centering
\end{minipage}
\caption{Fitted models and original dataset: females (left panel), males (right panel)}
\label{fig:TotnoRe}
\end{figure}

\begin{figure}[h!]
\hspace{-3pc}
\centering
\begin{minipage}{0.48\textwidth}
\includegraphics[width=8.8cm]{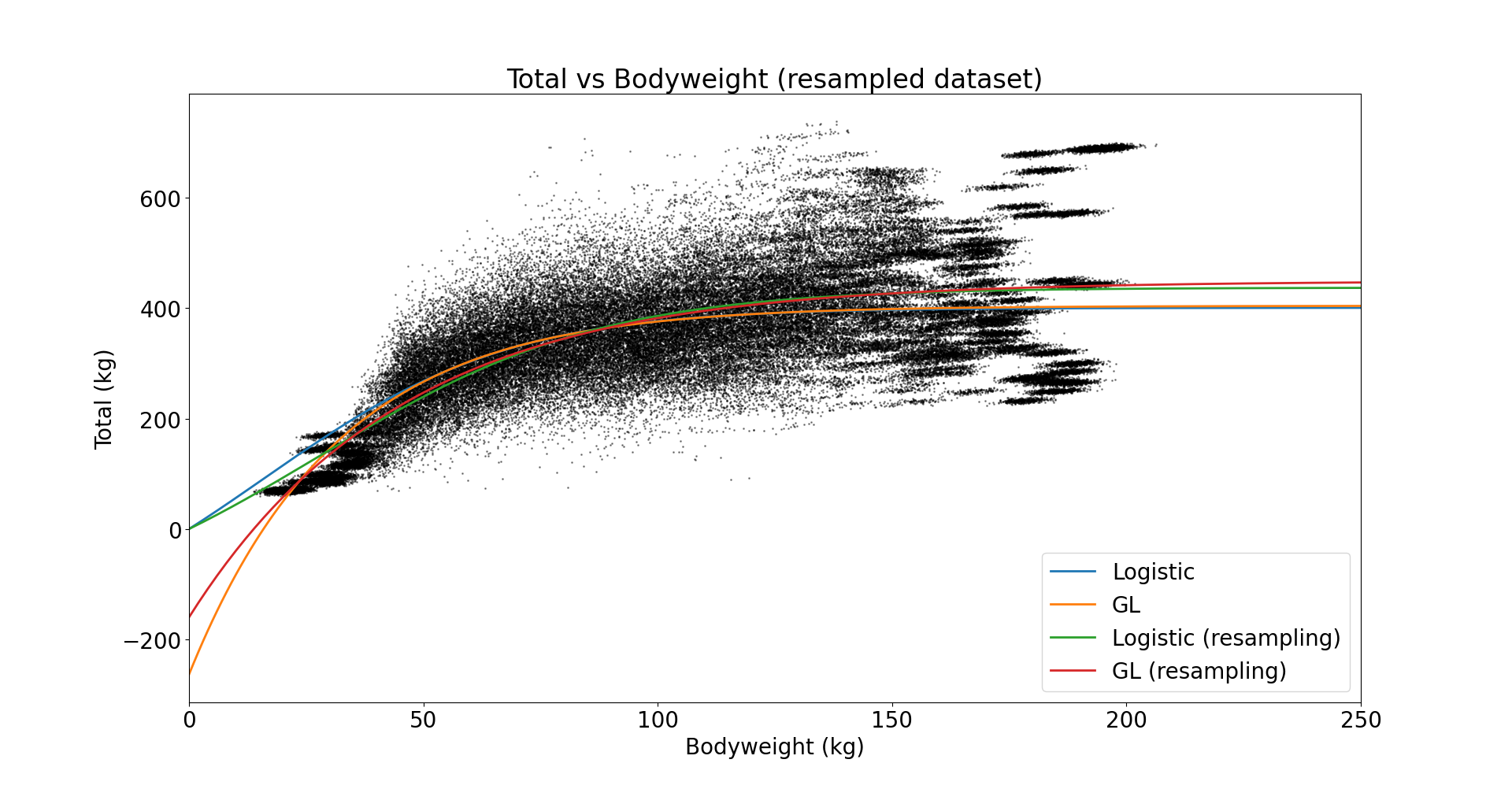}
\centering
\end{minipage}
\begin{minipage}{0.48\textwidth}
\includegraphics[width=8.8cm]{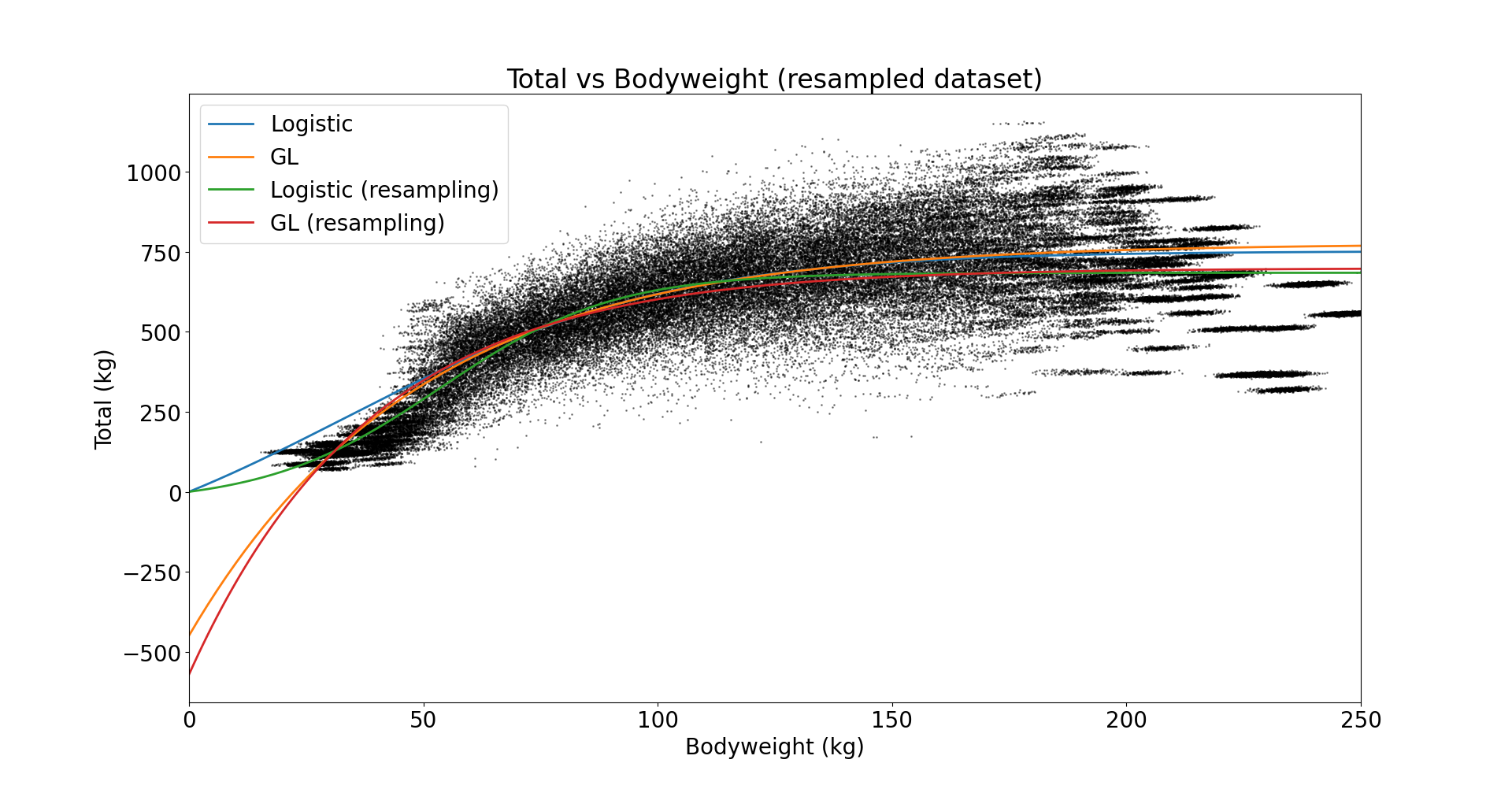}
\centering
\end{minipage}
\caption{Fitted models and resampled dataset: females (left panel), males (right panel)}
\label{fig:TotRe}
\end{figure}

Coefficients of the fitted are provided so readers can calculate scores -- see Tables \ref{tab:VBtable} and \ref{tab:LMtable} -- and possibly compare the resulting systems to other scoring methods. Plots are provided as a preliminary visual check in Figures \ref{fig:TotnoReMyr}, \ref{fig:TotnoRe} and \ref{fig:TotRe}. 
\begin{table}[h]
    \centering   
\begin{tabular}{ cc }   % top level tables, with 2 columns
\textbf{Females} & \textbf{Males} \\
\begin{tabular}{|c|c|c|c|}
\hline
         &$L (10^2kg)$ & $k(10^{-2}kg^{-1})$ & $x_0 (kg)$\\ \hline
        Original & $4.044$ & $3.811$ & $15.89$ \\ 
        Resampled & $4.505$ & $2.180$ & $13.63$ \\
        \hline
    \end{tabular} &  % starting rightmost sub table
% table 2
\begin{tabular}{|c|c|c|c|}
\hline
         &$L (10^2kg)$ & $k(10^{-2}kg^{-1})$ & $x_0 (kg)$\\\hline
        Original & $7.767$ & $2.045$ & $22.33$ \\ 
        Resampled & $6.994$ & $2.577$ & $23.14$ \\
        \hline
    \end{tabular} \\
\end{tabular}
\caption{Fitted coefficients for the Von Bertalanffy model}
    \label{tab:VBtable}
\end{table}
\vspace{-15pt}
\begin{table}[h]
\centering
\begin{tabular}{ cc }   % top level tables, with 2 columns
\textbf{Females} & \textbf{Males} \\
\begin{tabular}{|c|c|c|c|}
\hline
         &$L (10^2kg)$ & $k(10^{-2}kg^{-1})$ & $x_0 (kg)$\\ \hline
        Original & $6.189$ & $3.811$ & $15.97 $ \\ 
        Resampled & $6.308$ & $3.2019$ & $25.87 $ \\
\hline
    \end{tabular} &
\begin{tabular}{|c|c|c|c|}
\hline
         &$L (10^2kg)$ & $k(10^{-2}kg^{-1})$ & $x_0 (kg)$\\\hline
        Original & $10.39$ & $2.877$ & $33.51 $ \\ 
        Resampled& $7.223$ & $5.447$ & $53.40$ \\
        \hline
    \end{tabular}\\
\end{tabular}
\caption{Fitted coefficients for the logistic model}
    \label{tab:LMtable}
\end{table}
\vspace{-15pt}

\vspace{-20pt}
\vspace{-15pt}

\newpage
\subsection{Score distribution and homogeneity}
\begin{figure}[h!]
\hspace{-3pc}
\centering
\begin{minipage}{0.48\textwidth}
\includegraphics[width=8.8cm]{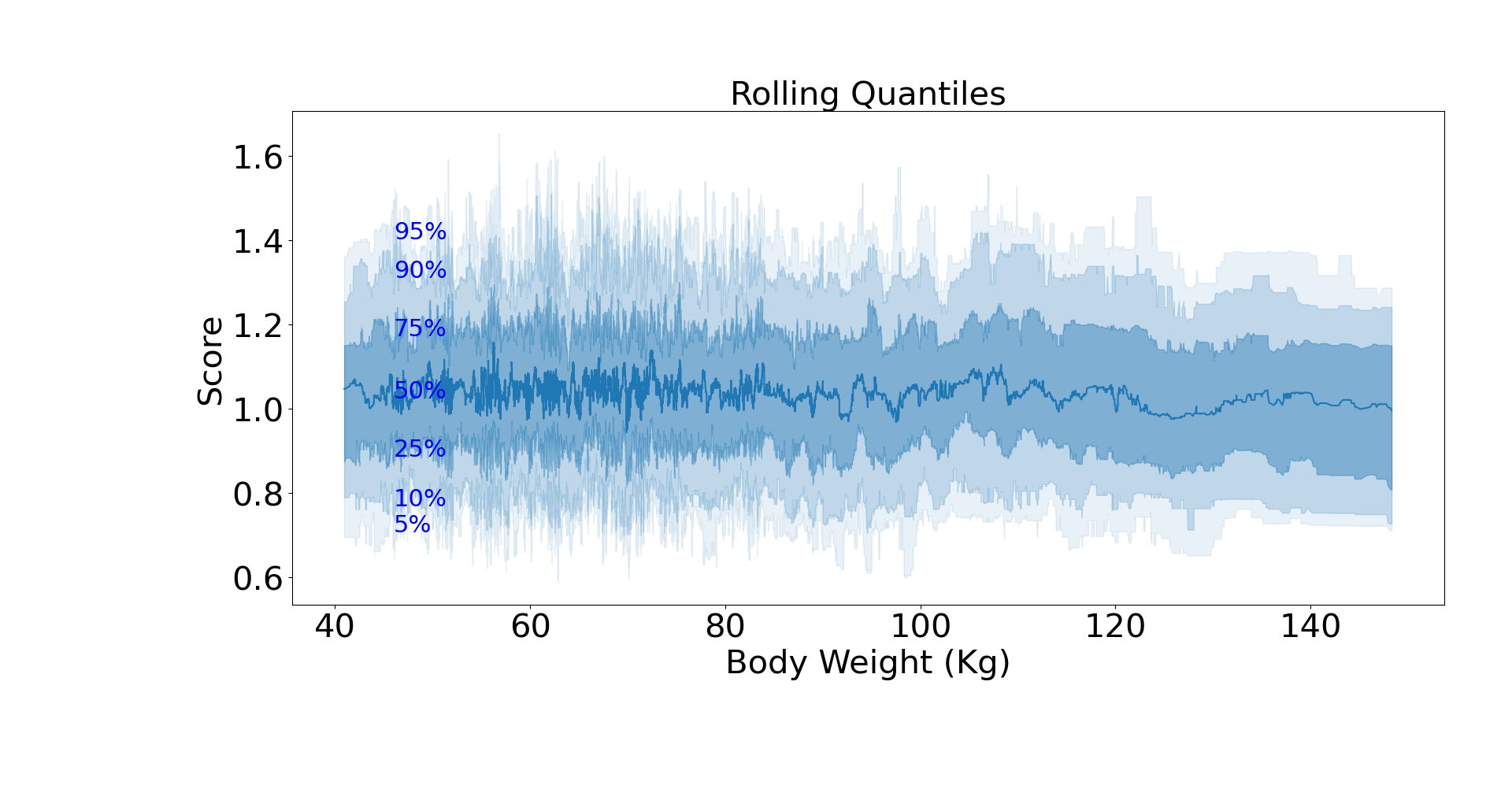}
\end{minipage}
\begin{minipage}{0.48\textwidth}
\includegraphics[width=8.8cm]{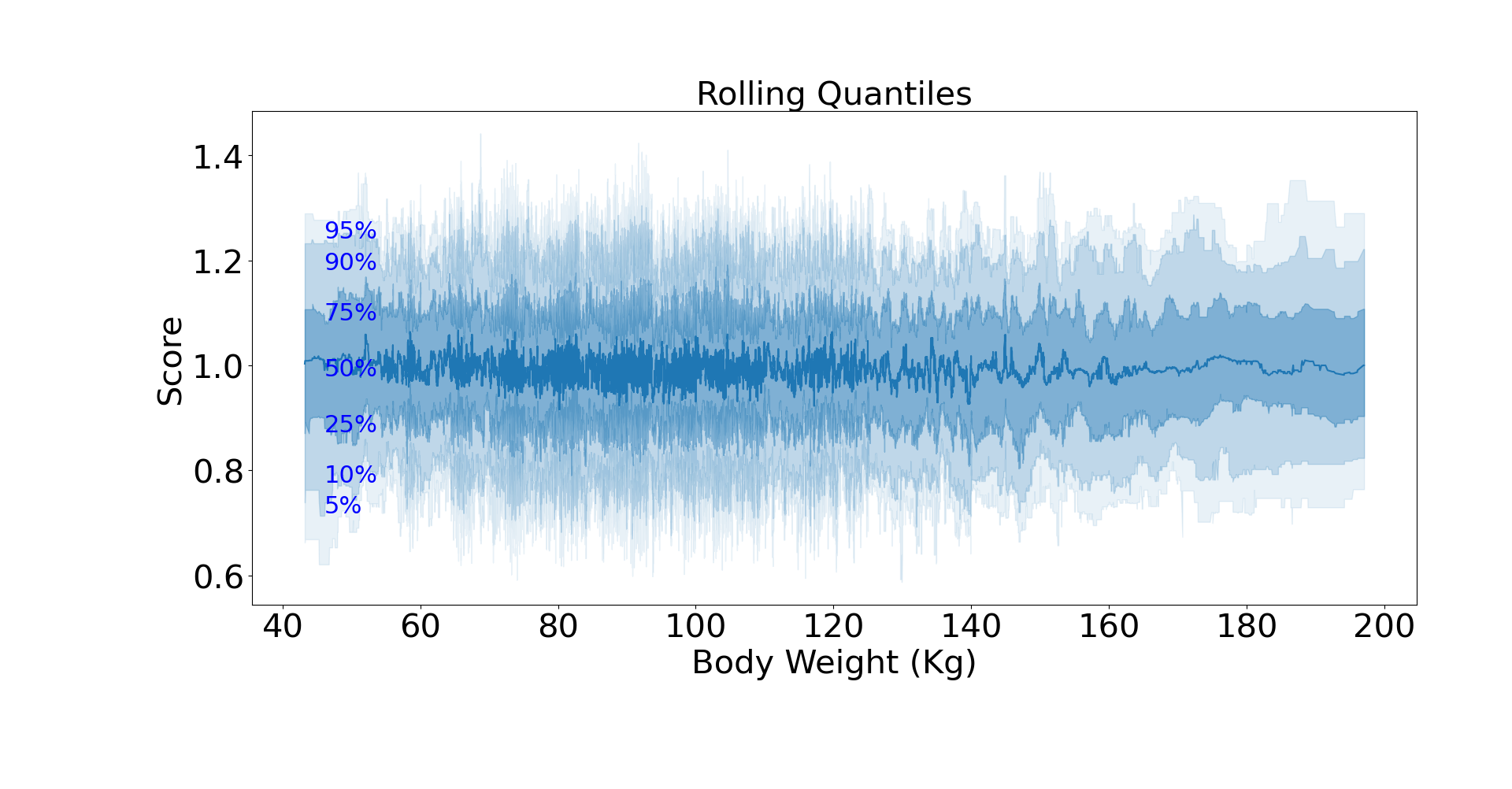}
\centering
\end{minipage}
\caption{Rolling Quantiles for the Score Distribution (Window size=$100$) : females (left panel), males (right panel)}
\label{fig:quantiles}
\end{figure}

\begin{figure}[h!]
\centering
\begin{minipage}{0.48\textwidth}
\includegraphics[width=8.8cm]{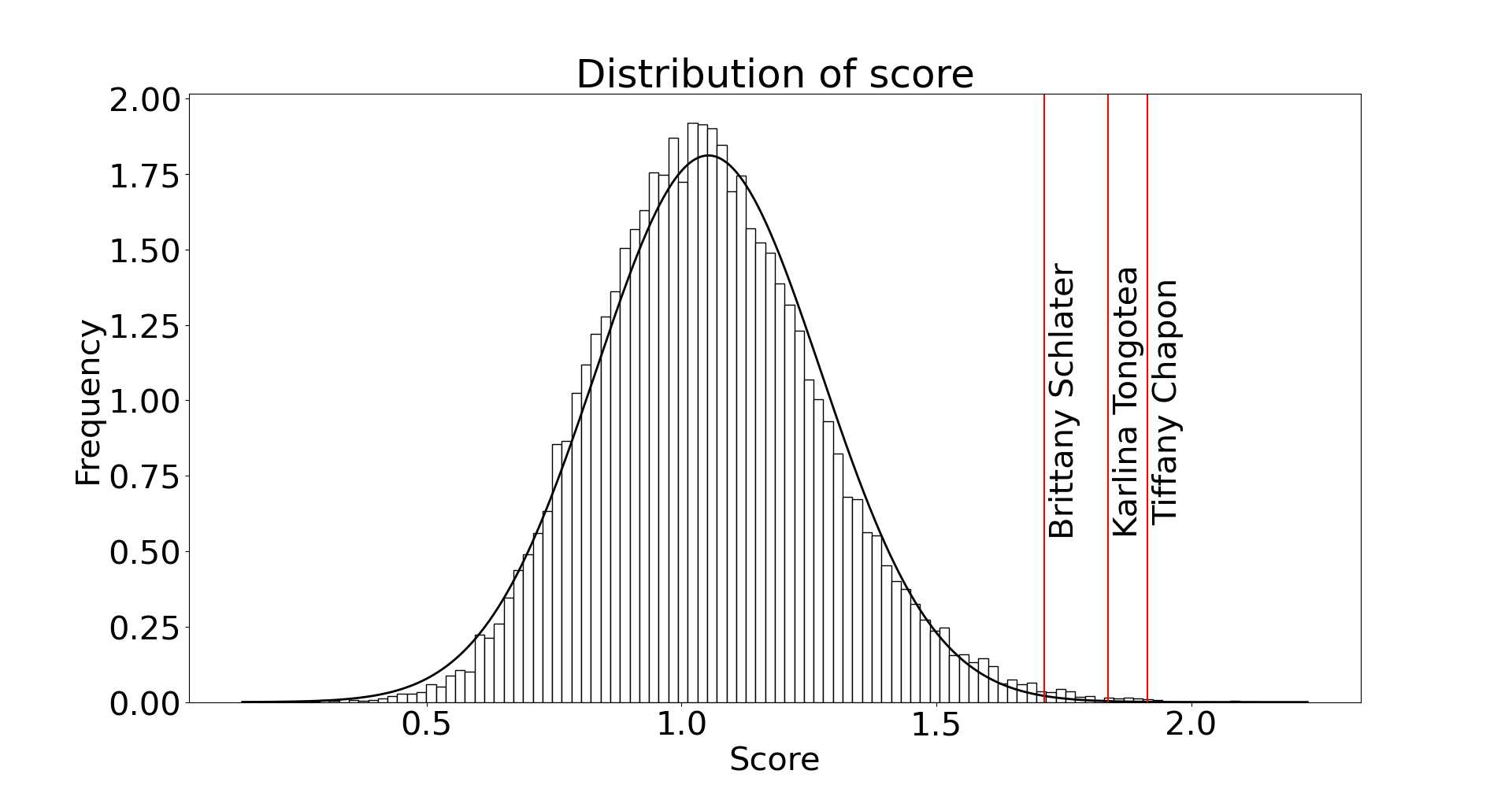}
\centering
\end{minipage}
\begin{minipage}{0.48\textwidth}
\includegraphics[width=8.8cm]{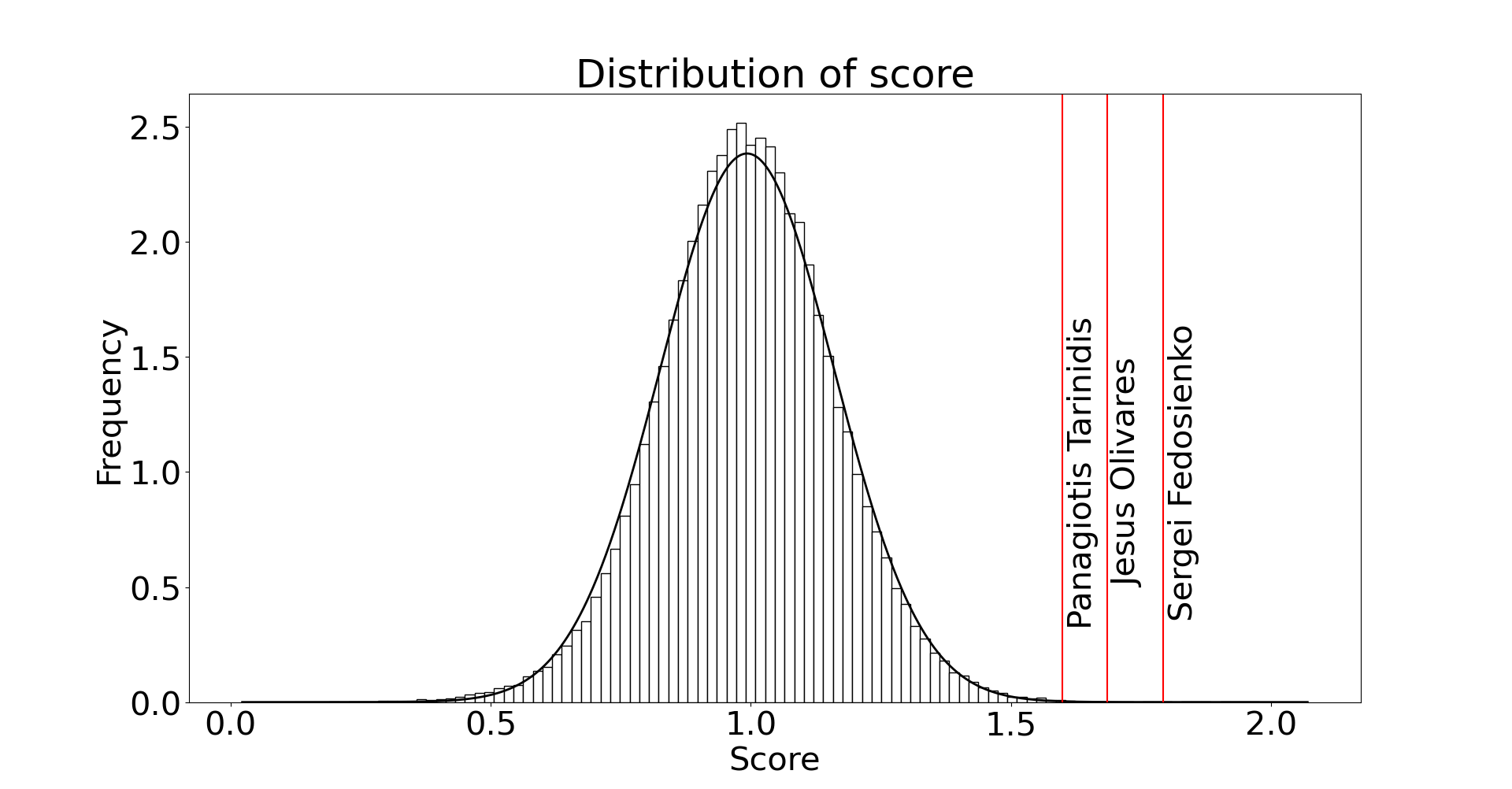}
\centering
\end{minipage}
\caption{Score Distribution and Gaussian Fit : females (left panel), males (right panel)}
\label{fig:ScoreDist}
\end{figure}

Figure \ref{fig:quantiles} shows that the quantiles are aligned through the body weight range and does not indicate any obvious bias. The distributional properties of the female and the male score appear to differ according to Figure \ref{fig:ScoreDist}. In particular, the females score appears to be more skewed than the males score. This suggests that this score may not be suitable to compare females and males.
\vspace{-5pt}
\vspace{-15pt}

\subsection{General observations}

The logistic model fitted with the resampled dataset locates an inflection point at $x_0=53.4\, kg$ for males. This can be considered as a low body weight, but it is sustainable for younger athletes, who will gain mass during their career, and for athletes of shorter stature. Men who weigh less than $53.4 \, kg$ represent  about $0.3\%$ of the sample. Although it is a low proportion, it is non-negligible, especially if we want to construct a score that would be applicable to a more general population— that is athletes who specialise in other sports or non-athletes. We should mention that this coincides with the $53 \, kg$ class, which is the lightest junior weight class in the IPF federation.

For females, the inflection point is located at $x_0= 23.87\, kg$. We count only $3$ athletes who are below this threshold, they represent less than $0.004 \%$ of the sample. This indicates that either the data used here is not sufficient to fit the logistic model, as the resampling method is forced to reproduce many instances of potential outliers that represent the lower body weight range, or that a more refined model is necessary.

We would need to collect data from non-competitors, should we consider to construct a scoring system to assess the strength level in a more general population. However, this collection would be challenging as it would require to ensure that movements are performed to a consistent standard. This rules out self-collected data and limits the possibilities for collecting a large amount of data.

\section*{Reproducibility and Data Availability}
The figures presented in this paper can be reproduced using the code available at the GitHub repository \cite{Repo}. The latest version of OpenPowerlifting dataset can be downloaded at \cite{openpower}.

\section*{Acknowledgment}
This research project was conducted in part during a visit from J.C.P., funded by the University of Leeds, starting on 16/12/24 and ending on 20/12/24.

\bibliography{references.bib}

\end{document}